\documentclass[a4paper,aps,prb,reprint,floatfix,amsmath,amssymb,amsfonts,longbibliography]{revtex4-1}
\usepackage{mathptmx}
\usepackage[scaled=0.9]{helvet}
\usepackage[utf8]{inputenc}
\usepackage{textcomp}
\usepackage{amsmath}
\usepackage{graphicx}
\usepackage{esint}

\makeatletter

\@ifundefined{textcolor}{}
{%
 \definecolor{BLACK}{gray}{0}
 \definecolor{WHITE}{gray}{1}
 \definecolor{RED}{rgb}{1,0,0}
 \definecolor{GREEN}{rgb}{0,1,0}
 \definecolor{BLUE}{rgb}{0,0,1}
 \definecolor{CYAN}{cmyk}{1,0,0,0}
 \definecolor{MAGENTA}{cmyk}{0,1,0,0}
 \definecolor{YELLOW}{cmyk}{0,0,1,0}
}

\usepackage{upgreek}% upright greek symbols
\usepackage[dvipsnames]{xcolor}
\usepackage{soul}% provides highlighting
\usepackage{xspace}
\usepackage[rightcaption]{sidecap}
\usepackage{epstopdf}
\usepackage{units}

\DeclareMathAlphabet{\mathcal}{OMS}{cmsy}{m}{n} %for nicer calligraphy

\setcitestyle{numbers,square}

\makeatother

\begin{document}

\title{Dislocation contrast in cathodoluminescence and electron-beam induced
current maps on GaN(0001)}

\author{Karl K. Sabelfeld}

\affiliation{Institute of Computational Mathematics and Mathematical Geophysics,
Russian Academy of Sciences and Novosibirsk State University, Lavrentiev
Prosp. 6, 630090 Novosibirsk, Russia.}

\author{Vladimir M. Kaganer}

\author{Carsten Pfüller}

\author{Oliver Brandt}

\affiliation{Paul-Drude-Institut für Festkörperelektronik, Hausvogteiplatz 5–7,
10117 Berlin, Germany}

\date{\today}
\begin{abstract}
We theoretically analyze the contrast observed at the outcrop of a
threading dislocation at the GaN(0001) surface in cathodoluminescence
and electron-beam induced current maps. We consider exciton diffusion
and recombination including finite recombination velocities both at
the planar surface and at the dislocation. Formulating the reciprocity
theorem for this general case enables us to provide a rigorous analytical
solution of this diffusion-recombination problem. The results of the
calculations are applied to an experimental example to determine both
the exciton diffusion length and the recombination strength of threading
dislocations in a free-standing GaN layer with a dislocation density
of $6\times10^{5}$~cm$^{-2}$. 
\end{abstract}
\maketitle

\section{Introduction}

Early GaN-based light emitting diodes exhibited an external quantum
efficiency of 4\% despite containing threading dislocations with a
density exceeding $10^{10}$~cm$^{-2}$ \cite{lester95}. This fact
was met with considerable surprise, since other III-V semiconductors
require drastically lower dislocation densities for reaching comparable
efficiencies. To explain this phenomenon, it was proposed that dislocations
in group-III nitrides are more benign than in other III-V compounds
due to their high ionicity \cite{lester95}. Alternatively, it was
suggested that the diffusion length of minority charge carriers in
GaN is smaller than the average distance between dislocations \cite{rosner97}.
In later work, however, it was often reported that the diffusion length
seems to be actually limited by this distance \cite{sugahara98,bandic00,chernyak01,karpov02}.

In cathodoluminescence (CL) or electron-beam induced current (EBIC)
maps of the surface of heteroepitaxial GaN(0001) layers, threading
dislocations are observed as dark spots directly reflecting the fact
that they represent nonradiative centers \cite{rosner97}. Intensity
profiles recorded across these areas of reduced luminous efficiency
contain, in principle, information on both the minority-carrier (or,
in GaN, exciton) diffusion length and the recombination strength of
the dislocation \cite{donolato79}. To determine these quantities,
however, it is required to consider the generation, diffusion and
recombination of electron-hole pairs in the presence of a surface
and a dislocation both possessing finite recombination strengths.
This problem belongs to the classical problems of semiconductor physics
and is under investigation since the 1950s \cite{haynes51,visvanathan54,mckelvey55,vanroosbroeck55},
but a general solution has not been obtained so far. In particular,
most of the existing theoretical work \cite{donolato78,donolato79,jakubowicz85,donolato85,jakubowicz86,donolato92,donolato98}
is restricted to either zero or infinite surface recombination velocity
in addition to an infinite recombination strength at the dislocation,
with the sole exception of the study by Jakubowicz \cite{jakubowicz86}
in which a finite recombination strength was introduced in an \emph{ad
hoc} fashion.

This body of theoretical understanding has been largely ignored by
experimentalists. Rosner \emph{et al}. \cite{rosner97} were the first
to correlate the outcrops of dislocations visible in atomic force
micrographs of a GaN(0001) layer with the contrast fluctuations in
CL maps. Due to the high dislocation density in the layer under investigation,
individual dislocations were not resolved in the CL map, and the authors
hence resorted to a simple phenomenological expression (first proposed
by Suzuki and Matsumoto \cite{suzuki75}) for describing the impact
of dislocations on the CL intensity. In subsequent experimental work,
individual dislocations were resolved in CL maps which would have
been eligible objects for a more detailed analysis using appropriate
models. However, in all of these studies \cite{rosner97,sugahara98,chernyak01,cherns01,nakaij05,pauc06,ino08,pauc10},
the simple expression proposed by Rosner \emph{et al}.\cite{rosner97}
was adopted despite the fact that it does not represent a sensible
approximation of the intensity profile across threading dislocations
even in the most simple case of zero surface recombination velocity
and infinite recombination strength at the dislocations.

This fact has recently been noted by Yakimov \cite{yakimov10}, who
pointed out that the analysis of the experimental data in this purely
empirical way leads to a significant underestimation of the exciton
diffusion length. As a consequence, values for the exciton diffusion
length derived in this way from CL or EBIC maps are to be viewed with
caution \cite{yakimov15}. Since either of these techniques gives,
in principle, simultaneous access to three technologically important
quantities (the exciton diffusion length, the surface recombination
velocity, and the dislocation recombination strength), the problem
clearly merits further theoretical and experimental attention.

In the present paper, we derive a rigorous analytical solution for
the problem of excitons being generated within a finite volume and
subsequently diffusing and recombining in the bulk as well as annihilating
at the surface and at threading dislocations with arbitrary recombination
velocities. For this purpose, the reciprocity theorem \cite{donolato85}
is generalized. The solution of the problem for finite exciton recombination
velocities at the planar surface is obtained by considering, instead
of the half-space, a finite-thickness slab, whose thickness can be
taken sufficiently large. This solution is also generalized for the
case of thin films with finite recombination velocity at the back
surface, as it is required for solar cells \cite{kieliba06,budhraja13}.
The solution for a point source of excitons (the Green function) is
directly convolved with the numerically calculated distributions of
excitons generated by electron beams of different energies. The results
of the calculations are applied to describe the contrast of threading
dislocations in a freestanding GaN layer with an average distance
between dislocations of more than 10 \textmu m. A fit of the expression
derived in this work to an experimental profile yields both the exciton
diffusion length and the recombination strength at the dislocation.

\section{General equations\label{sec:equations}}

\begin{figure}
\includegraphics[width=0.75\columnwidth]{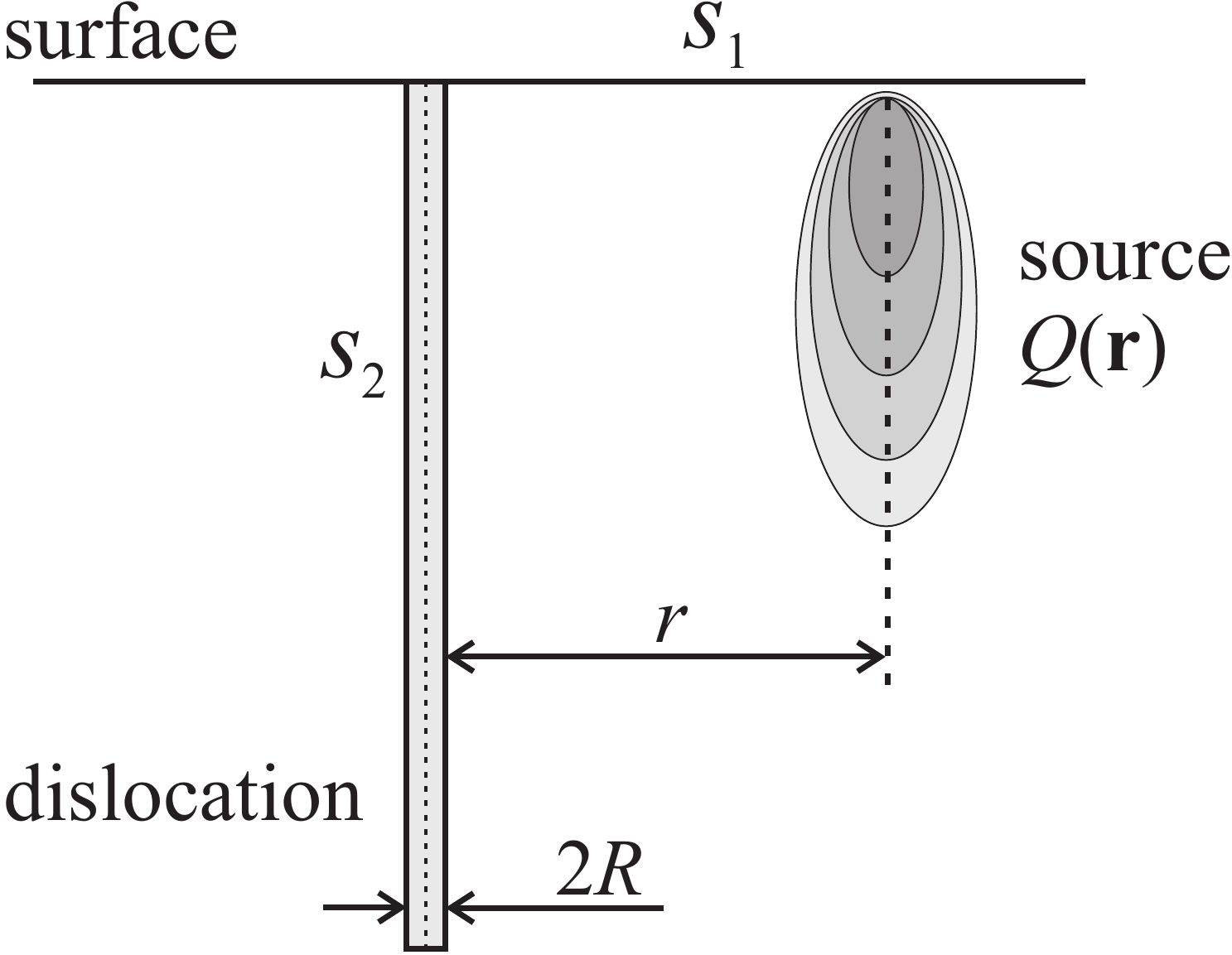}

\protect\protect\caption{Schematic depiction of a CL or EBIC experiment in the close vicinity
of a surface and a dislocation. The incident electron beam creates
a spatial distribution $Q(\mathbf{r})$ of excitons, which then diffuse
with a diffusivity $D$ and recombine with an effective lifetime $\tau$
in a semi-infinite layer with a surface having a surface recombination
velocity $S_{1}$ intersected by a threading dislocation with a recombination
velocity $S_{2}$.}

\label{fig:sketch} 
\end{figure}

We consider the generation of electron-hole pairs by an electron beam
incident on the surface of a GaN\{0001\} layer with the three-dimensional
generation function $Q(\mathbf{r})$. Because of the exciton binding
energy of 26\,meV in GaN, these electron-hole pairs rapidly bind
to form excitons which then diffuse as a stable entity up to room
temperature. Our aim is to solve the exciton diffusion-recombination
problem in the half-space containing a dislocation line normal to
the surface, as shown in Fig.~\ref{fig:sketch}. After forming from
the high energy electron-hole cloud generated by the incident electron
beam, these excitons will diffuse, recombine in the bulk, or annihilate
nonradiatively at both the free surface of the half-space and the
dislocation.

The diffusion and recombination of excitons generated by the source
$Q(\mathbf{r})$ are described by the three-dimensional diffusion-recombination
equation 
\begin{equation}
D\Delta n(\mathbf{r})-n(\mathbf{r})/\tau+Q(\mathbf{r})=0,\label{eq:1}
\end{equation}
where $D$ is the diffusion coefficient and $\tau$ the lifetime of
excitons, describing their recombination in the bulk. This latter
quantity includes both radiative and nonradiative contributions, i.\,e.,
$1/\tau=1/\tau_{r}+1/\tau_{nr}$.

Nonradiative exciton annihilation at the surfaces is described by
the boundary conditions 
\begin{equation}
D\frac{\partial n(\mathbf{r})}{\partial\nu_{k}}+S_{k}n(\mathbf{r})=0,\,\,\,\mathbf{r}\in\Gamma_{k},\label{eq:2}
\end{equation}
where $\Gamma_{k}$ are the surfaces, $\nu_{k}$ the surface normals,
$S_{k}$ the surface recombination velocities, and $k=1,2$ correspond
to the planar surface of the half-space and the cylindrical surface
of the dislocation, respectively.

The total number of excitons annihilating at the respective surfaces
is equal to the fluxes integrated over these surfaces, 
\begin{equation}
I_{k}=D\int_{\Gamma_{k}}\frac{\partial n(\mathbf{r})}{\partial\nu_{k}}d\sigma_{k}.\label{eq:3}
\end{equation}
Particularly, the flux of excitons to the planar surface $I_{1}$
can be measured in an EBIC experiment, using a Schottky barrier to
dissociate excitons into electrons and holes and separating them prior
to recombination. The CL intensity is proportional to the total number
of excitons recombining radiatively in the bulk. This quantity can
be calculated by subtracting the fluxes to the surfaces $I_{1}$ and
$I_{2}$ from the total number of excitons produced by the source,
$I_{0}=\int Q(\mathbf{r})dV$, so that $I_{\mathrm{CL}}=\eta(I_{0}-I_{1}-I_{2})$,
where $\eta=\tau/\tau_{r}$ is the internal quantum efficiency, i.\,e.,
the fraction of excitons recombining radiatively. Since the measured
CL intensity is arbitrarily scaled, the factor $\eta$ can be included
in the normalization constant and is omitted in the further analysis.
Thus, our purpose is to calculate the fluxes to the surfaces $I_{1}$
and $I_{2}$.

It is convenient to multiply Eqs.~(\ref{eq:1})–(\ref{eq:3}) with
$\tau$ and to introduce the diffusion length $\Lambda=(D\tau)^{1/2}$
and the surface recombination lengths $l_{k}=S_{k}\tau$. The length
$l_{1}$ describes recombination at the planar surface, while the
length $l_{2}$ is the corresponding quantity at the dislocation.
Then, all parameters entering the problem are quantities with the
dimension of lengths: in addition to the three lengths $\Lambda,\, l_{1},\, l_{2}$,
we need to consider the radius $R$ of the dislocation. Since Eqs.~(\ref{eq:1})–(\ref{eq:3})
are linear, it is sufficient to consider a point source of excitons
in an arbitrary point $\mathbf{r}'$, $Q(\mathbf{r})=\delta(\mathbf{r}-\mathbf{r}')$,
and calculate the Green function $G(\mathbf{r},\mathbf{r}')$. Then,
the solution for any finite source $Q(\mathbf{r})$ can be obtained
as a convolution 
\begin{equation}
I(\mathbf{r})=\int G(\mathbf{r},\mathbf{r}')Q(\mathbf{r}')d\mathbf{r}'.\label{eq:3a}
\end{equation}

Hence, the problem can be finally formulated as follows. The distribution
of excitons is governed by the three-dimensional diffusion-annihilation
equation 
\begin{equation}
\Lambda^{2}\Delta G(\mathbf{r},\mathbf{r}')-G(\mathbf{r},\mathbf{r}')+\delta(\mathbf{r}-\mathbf{r}')=0,\label{eq:4}
\end{equation}
with third-type boundary conditions (Robin boundary conditions) at
the planar surface of half-space ($k=1$) and at the cylindrical surface
of the dislocation ($k=2$) 
\begin{equation}
\Lambda^{2}\frac{\partial G(\mathbf{r},\mathbf{r}')}{\partial\nu_{k}}+l_{k}G(\mathbf{r},\mathbf{r}')=0,\,\,\,\mathbf{r}\in\Gamma_{k}.\label{eq:5}
\end{equation}
The boundary condition at large distances from the source and the
dislocation is $G(\mathbf{r},\mathbf{r}')\rightarrow0$ at $\mathbf{r}\rightarrow\infty$.
Our aim is to calculate the total fluxes to each surface, 
\begin{equation}
I_{k}(\mathbf{r}')=\int_{\Gamma_{k}}\frac{\partial G(\mathbf{r},\mathbf{r}')}{\partial\nu_{k}}d\sigma_{k}.\label{eq:6}
\end{equation}

\section{The reciprocity theorem\label{sec:Reciprocity}}

The reciprocity theorem was formulated and proven by Donolato \cite{donolato85}
for the case when the planar surface is a perfect sink (infinite surface
recombination velocity): in this case, the local generation of excitons
with integral detection of their recombination is equivalent to their
uniform generation with local detection. Here, we extend this theorem
to arbitrary finite surface recombination velocities.

Let us show that the integral $I_{i}(\mathbf{r})$ given by Eq.~(\ref{eq:6})
is identically equal to $\Lambda^{2}F(\mathbf{r})$, where $F(\mathbf{r})$
is the solution of the following homogeneous boundary value problem:
\begin{equation}
\Lambda^{2}\Delta F(\mathbf{r})-F(\mathbf{r})=0,\label{eq:7}
\end{equation}
with the boundary condition at the $i$th surface modified as follows:
\begin{equation}
\Lambda^{2}\frac{\partial F(\mathbf{r})}{\partial\nu_{i}}+l_{i}[F(\mathbf{r})-1]=0,\,\,\,\mathbf{r}\in\Gamma_{i},\label{eq:8}
\end{equation}
while on all other surfaces ($k\neq i$) the boundary condition remains
unchanged, 
\begin{equation}
\Lambda^{2}\frac{\partial F(\mathbf{r})}{\partial\nu_{k}}+l_{k}F(\mathbf{r})=0,\,\,\,\mathbf{r}\in\Gamma_{k},\, k\neq i.\label{eq:9}
\end{equation}

The proof is based on Green's formula 
\begin{equation}
\int(F\Delta G-G\Delta F)dV=\sum_{k}\int_{\Gamma_{k}}\left(F\frac{\partial G}{\partial\nu_{k}}-G\frac{\partial F}{\partial\nu_{k}}\right)d\sigma_{k},\label{eq:10}
\end{equation}
where the integral in the left-hand side is over the volume and the
integral in the right-hand side is a sum of integrals over all surfaces
bounding this volume. A substitution of Eqs.~(\ref{eq:4}) and (\ref{eq:7})
reduces Green's formula to 
\begin{equation}
\frac{F(\mathbf{r}')}{\Lambda^{2}}=-\sum_{k}\int_{\Gamma_{k}}\left(F(\mathbf{r})\frac{\partial G(\mathbf{r},\mathbf{r}')}{\partial\nu_{k}}-G(\mathbf{r},\mathbf{r}')\frac{\partial F(\mathbf{r})}{\partial\nu_{k}}\right)d\sigma_{k}.\label{eq:11}
\end{equation}
All surface integrals with $k\neq i$ vanish since the functions $F$
and $G$ satisfy the same boundary conditions, (\ref{eq:5}) and (\ref{eq:9}),
at these surfaces. The integral over the remaining boundary $\Gamma_{i}$
coincides with the right-hand side of Eq.~(\ref{eq:6}), which completes
the proof.

\section{Calculation of the fluxes\label{sec:Calculation}}

Before considering the most general case of finite recombination velocities
on both surfaces, the planar surface of the half-space and the cylindrical
surface of the dislocation, we analyze two limiting cases. In the
first case, we assume that excitons do not annihilate at the planar
surface ($l_{1}\rightarrow0$). An exciton diffusing to the surface
is thus “reflected” by it and continues its diffusional motion in
the half-space. The continuation of its random trajectory is a mirror
reflection of the trajectory that it would have in an infinite space
with an infinitely long dislocation. Thus, in this case the solution
of the diffusion problem in the half-space is the same as in the infinite
space.

The second case is the one considered by Donolato \cite{donolato98}:
the planar surface is a perfect sink for excitons ($l_{1}\rightarrow\infty$).
In this case we calculate, in addition to the total flux of excitons
to the planar surface, the total flux to the dislocation. The former
quantity (calculated by Donolato) gives the EBIC intensity (requiring
a Schottky barrier or a \emph{p-n} junction to dissociate the excitons),
while the latter is needed to obtain the CL intensity.

\subsection{Reflection (Neumann) boundary condition at the planar surface}

\label{subsec:refl}

In this case ($l_{1}\rightarrow0$) there is no flux at the planar
surface, and we need to calculate the flux to the cylindrical surface
of the dislocation. From the reciprocity theorem, this flux $f(r,z)$
(where $r,z$ are cylindrical coordinates of the point source of excitons)
is a solution of the equation 
\begin{equation}
\Delta_{r}f(r,z)+\frac{\partial^{2}f(r,z)}{\partial z^{2}}-\Lambda^{-2}f(r,z)=0\label{eq:12}
\end{equation}
with the boundary conditions 
\begin{equation}
\left.\frac{\partial f}{\partial z}\right|_{z=0}=0,\,\,\,\,\left.\left(-\Lambda^{2}\frac{\partial f}{\partial r}+l_{2}f\right)\right|_{r=R}=l_{2}.\label{eq:13}
\end{equation}
The derivative $\partial f/\partial r$ enters the second boundary
condition (\ref{eq:13})$_{2}$ with negative sign, since the outer
normal $\nu_{2}$ in Eqs.~(\ref{eq:10}) and (\ref{eq:11}) is along
the negative direction of $r$.

Since the reflecting planar boundary can be replaced by continuation
of the medium to the other half-space, we seek for a solution that
does not depend on $z$. Equation (\ref{eq:12}) and the first boundary
condition (\ref{eq:13})$_{1}$ are satisfied with $f(r,z)=CK_{0}(r/\Lambda)$,
where $K_{0}(x)$ is the zero order modified Bessel function of the
second kind. The coefficient $C$ is determined from the second boundary
condition (\ref{eq:13})$_{2}$. The solution is 
\begin{equation}
f(r)=\frac{l_{2}K_{0}(r/\Lambda)}{\Lambda K_{1}(R/\Lambda)+l_{2}K_{0}(R/\Lambda)},\label{eq:14}
\end{equation}
where $K_{1}(x)$ is the first order modified Bessel function of the
second kind.

For the discussion below in Sec.~\ref{subsec:Finite}, it is instructive
to present also a more formal, albeit more lengthy, derivation of
Eq.~(\ref{eq:14}). A solution of Eq.~(\ref{eq:12}) can be sought
in the form 
\begin{equation}
f(r,z)=\intop_{0}^{\infty}C(q)K_{0}(\mu_{q}r)\cos(qz)\, dq,\label{eq:15}
\end{equation}
where 
\begin{equation}
\mu_{q}=\sqrt{q^{2}+\Lambda^{-2}}.\label{eq:16}
\end{equation}
The term $\cos(qz)$ in the integral provides the first boundary condition
(\ref{eq:13})$_{1}$. The second boundary condition (\ref{eq:13})$_{2}$
can be written as 
\begin{equation}
\intop_{0}^{\infty}C(q)\left[\Lambda K_{1}(R/\Lambda)+l_{2}K_{0}(R/\Lambda)\right]\cos(qz)\, dq=l_{2}.\label{eq:16a}
\end{equation}
For the right-hand side of this equation, we use the identity 
\begin{equation}
1=2\intop_{0}^{\infty}\delta(z)\cos(qz)\, dz,\label{eq:16b}
\end{equation}
where $\delta(x)$ is the Dirac delta function, and thus obtain

\begin{equation}
C(q)=\frac{2l_{2}\delta(q)}{\Lambda K_{1}(R/\Lambda)+l_{2}K_{0}(R/\Lambda)}.\label{eq:17}
\end{equation}
Substituting this expression into Eq.~(\ref{eq:15}), we arrive at
the same result (\ref{eq:14}).

\subsection{Absorption (Dirichlet) boundary condition at the planar surface}

\subsubsection{Flux to the dislocation}

We now consider the planar boundary as a perfect sink for excitons
with surface recombination velocity $l_{1}\rightarrow\infty$, and
calculate first the flux $f(x,z)$ of the excitons produced by a point
source at ($r,z$) to the cylinder $r=R$. The reciprocity theorem
provides for the function $f(r,z)$ the boundary value problem 
\begin{equation}
\Delta_{r}f(r,z)+\frac{\partial^{2}f(r,z)}{\partial z^{2}}-\Lambda^{-2}f(r,z)=0,\label{eq:26}
\end{equation}
\begin{equation}
\left.f\right|_{z=0}=0,\,\,\,\,\left.\left(-\Lambda^{2}\frac{\partial f}{\partial r}+l_{2}f)\right)\right|_{r=R}=l_{2}.\label{eq:27}
\end{equation}
We seek the solution in the form 
\begin{equation}
f(r,z)=\intop_{0}^{\infty}C(q)K_{0}(\mu_{q}r)\sin(qz)\, dq\label{eq:28}
\end{equation}
and represent the boundary condition at the cylinder surface (\ref{eq:27})$_{2}$
as 
\begin{equation}
\intop_{0}^{\infty}C(q)\left[l_{2}K_{0}(\mu_{q}R)+\Lambda^{2}\mu_{q}K_{1}(\mu_{q}R)\right]\sin(qz)\, dq=l_{2}.\label{eq:29}
\end{equation}
Using in the right-hand side of this equation the equality 
\begin{equation}
1=\frac{2}{\pi}\intop_{0}^{\infty}\frac{\sin(qz)}{q}\, dq,\label{eq:30}
\end{equation}
we obtain the flux to the dislocation 
\begin{equation}
f(r,z)=\frac{2}{\pi}\intop_{0}^{\infty}\frac{l_{2}K_{0}(\mu_{q}r)\sin(qz)\, dq}{q[l_{2}K_{0}(\mu_{q}R)+\Lambda^{2}\mu_{q}K_{1}(\mu_{q}R)]}.\label{eq:31}
\end{equation}

\subsubsection{Flux to the planar boundary}

\label{subsec:AbsorbPlanar}

Let us calculate now the total flux $g(r,z)$ to the planar surface
of the excitons produced by a point source at ($x,z$). According
to the reciprocity theorem in Sec.~\ref{sec:Reciprocity}, the function
$g(r,z)$ satisfies the equation 
\begin{equation}
\Delta_{r}g(r,z)+\frac{\partial^{2}g(r,z)}{\partial z^{2}}-\Lambda^{-2}g(r,z)=0\label{eq:18}
\end{equation}
and the boundary conditions 
\begin{equation}
\left.g\right|_{z=0}=1,\,\,\,\,\left.\left(-\Lambda^{2}\frac{\partial g}{\partial r}+l_{2}g)\right)\right|_{r=R}=0.\label{eq:19}
\end{equation}

Let us find a function $v(z)$ satisfying Eq.~(\ref{eq:18}) and
the first boundary condition (\ref{eq:19})$_{1}$, i.e., 
\begin{equation}
d^{2}v/dz^{2}-\Lambda^{-2}v=0,\,\,\,\, v(0)=1.\label{eq:19a}
\end{equation}
The solution is $v(z)=\exp(-z/\Lambda)$. Let us introduce a new function
$u(r,z)$ by $g(r,z)=v(z)-u(r,z)$. Then, the boundary value problem
can be written as 
\begin{equation}
\Delta_{r}u(r,z)+\frac{\partial^{2}u(r,z)}{\partial z^{2}}-\Lambda^{-2}u(r,z)=0,\label{eq:20}
\end{equation}
\begin{equation}
\left.u\right|_{z=0}=0,\,\,\,\,\left.\left(-\Lambda^{2}\frac{\partial u}{\partial r}+l_{2}u)\right)\right|_{r=R}=l_{2}\exp(-z/\Lambda).\label{eq:21}
\end{equation}

We seek for a solution in the form 
\begin{equation}
u(r,z)=\intop_{0}^{\infty}C(q)K_{0}(\mu_{q}r)\sin(qz)\, dq.\label{eq:22}
\end{equation}
Equation (\ref{eq:20}) and the first boundary condition (\ref{eq:21})$_{1}$
are satisfied, and the boundary condition (\ref{eq:21})$_{2}$ at
the surface of the cylinder $r=R$ reads 
\begin{multline}
\intop_{0}^{\infty}C(q)\left[l_{2}K_{0}(\mu_{q}R)+\Lambda^{2}\mu_{q}K_{1}(\mu_{q}R)\right]\sin(qz)\, dq\\
=l_{2}\exp(-z/\Lambda).\label{eq:23}
\end{multline}
The right-hand side of this equation can be represented as 
\begin{equation}
l_{2}\exp(-z/\Lambda)=\frac{2}{\pi}l_{2}\intop_{0}^{\infty}\frac{q\sin(qz)}{q^{2}+\Lambda^{-2}}\, dq,\label{eq:24}
\end{equation}
which gives $C(q)$ and finally the flux to the planar surface 
\begin{multline}
g(r,z)=\exp(-z/\Lambda)\\
-\frac{2}{\pi}\intop_{0}^{\infty}\frac{ql_{2}}{q^{2}+\Lambda^{-2}}\,\frac{K_{0}(\mu_{q}r)\sin(qz)\, dq}{l_{2}K_{0}(\mu_{q}R)+\Lambda^{2}\mu_{q}K_{1}(\mu_{q}R)}.\label{eq:25}
\end{multline}

\subsection{Finite recombination velocity at the planar surface (Robin boundary
condition) \label{subsec:Finite}}

We have considered above two limiting cases, the absence of recombination
at the planar surface ($l_{1}\rightarrow0$) and the total annihilation
of excitons at this boundary ($l_{1}\rightarrow\infty$). In this
section, our aim is to generalize these results to the case of an
arbitrary surface recombination velocity (an arbitrary $l_{1}$).
For this purpose, we consider, instead of the half-space, a slab with
a finite thickness $t$, with the dislocation running across it. The
solution is represented by series, rather than integrals. Taking $t$
sufficiently large, we arrive at the solution for the half-space.
First, we assume the absence of surface recombination at the back
surface $z=t$. Second, we allow for an arbitrary surface recombination
velocity also at the back surface to obtain a solution for thin films
applicable, in particular, to the case of solar cells \cite{kieliba06,budhraja13}.

\subsubsection{Flux to the dislocation}

\begin{figure}
\includegraphics[width=0.7\columnwidth]{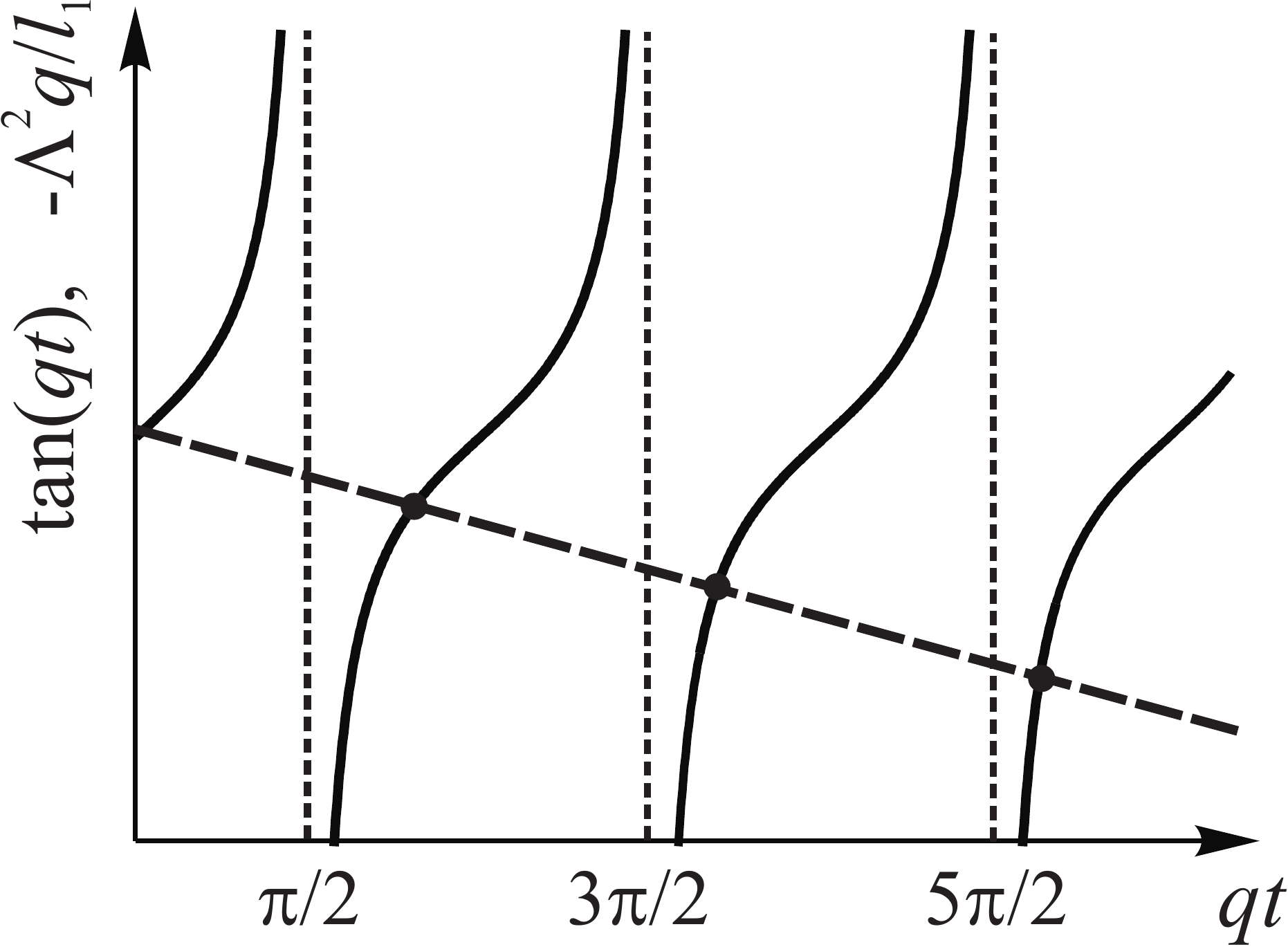}

\protect\protect\caption{Graphical representation of the eigenvalues $q_{n}$ of Eq.~(\ref{eq:37}). }

\label{fig:tan} 
\end{figure}

The reciprocity theorem states that the total flux $f(r,z)$ to the
surface of the cylinder from a point source at $(x,z)$ is a solution
of the diffusion equation 
\begin{equation}
\Delta_{r}f(r,z)+\frac{\partial^{2}f(r,z)}{\partial z^{2}}-\Lambda^{-2}f(r,z)=0\label{eq:32}
\end{equation}
with the boundary conditions 
\begin{gather}
\left.\left(-\Lambda^{2}\frac{\partial f}{\partial z}+l_{1}f\right)\right|_{z=0}=0,\,\,\,\left.\left(-\Lambda^{2}\frac{\partial f}{\partial r}+l_{2}f\right)\right|_{r=R}=l_{2},\nonumber \\
f(r,t)=0.\label{eq:33}
\end{gather}

The problem is solved by separation of vertical and radial variables.
We first seek for eigenfunctions $y_{n}(z)$ and eigenvalues $q_{n}$
($n=1,2,\ldots$) which are solutions of the ordinary differential
equation 
\begin{equation}
\frac{d^{2}y_{n}}{dz^{2}}+q_{n}^{2}y_{n}=0\label{eq:34}
\end{equation}
with the boundary conditions 
\begin{equation}
-\Lambda^{2}\left.\frac{dy_{n}}{dz}\right|_{z=0}+l_{1}y_{n}(0)=0,\,\,\, y_{n}(t)=0.\label{eq:35}
\end{equation}
The solutions of Eq.~(\ref{eq:34}) satisfying the second boundary
condition (\ref{eq:35})$_{2}$ are 
\begin{equation}
y_{n}(z)=p_{n}\sin\left[q_{n}(t-z)\right],\label{eq:36}
\end{equation}
where the coefficients $p_{n}$ will be chosen later. The first boundary
condition (\ref{eq:36})$_{1}$ gives rise to the equation 
\begin{equation}
\tan(q_{n}t)=-\Lambda^{2}q_{n}/l_{1}.\label{eq:37}
\end{equation}
This equation has an infinite number of solutions $q_{n}>0$, $n=1,2,\ldots$,
as illustrated in Fig.~\ref{fig:tan}. These solutions are the eigenvalues
of the problem. Evidently, the $n$th root is restricted by $n\pi/2<q_{n}t<\pi+n\pi/2$,
so that all roots can easily be determined numerically.

The limiting cases considered in the two previous sections can be
obtained from Eq.~(\ref{eq:37}). The reflection boundary condition
at the planar surface $l_{1}\rightarrow0$ gives $q_{n}t=(2n-1)\pi/2$,
and the eigenfunctions (\ref{eq:36}) are $y_{n}=p_{n}\cos(q_{n}z)$,
as also seen in Eq.~(\ref{eq:15}). In the opposite limit of the
absorption boundary condition, $l_{1}\rightarrow\infty$, the eigenvalues
are $q_{n}t=n\pi$, and the eigenfunctions $y_{n}=p_{n}\sin(q_{n}z)$,
as in Eqs.~(\ref{eq:28}) and (\ref{eq:22}).

The functions $y_{n}(z)$ are orthogonal on $[0,t]$, and we choose
the coefficients $p_{n}$ in Eq.~(\ref{eq:36}) by requiring the
functions to be normalized: 
\begin{equation}
\intop_{0}^{t}y_{n}(z)y_{n'}(z)\, dz=\delta_{nn'}.\label{eq:38}
\end{equation}
Making use of Eqs.~(\ref{eq:36}) and (\ref{eq:37}), we find 
\begin{equation}
p_{n}^{2}=2\left[t+\frac{l_{1}}{\left(l_{1}/\Lambda\right)^{2}+(\Lambda q_{n})^{2}}\right]^{-1}.\label{eq:39}
\end{equation}

Now we seek the solution in the form 
\begin{equation}
f(r,z)=\sum_{n=1}^{\infty}C_{n}y_{n}(z)K_{0}(\mu_{n}r),\label{eq:40}
\end{equation}
with 
\begin{equation}
\mu_{n}=\sqrt{q_{n}^{2}+\Lambda^{-2}},\label{eq:41}
\end{equation}
thus generalizing Eqs.~(\ref{eq:15}) and (\ref{eq:16}) to the case
of a finite slab thickness.

The boundary condition at the cylinder surface (\ref{eq:33})$_{2}$
reads 
\begin{equation}
\sum_{n=1}^{\infty}C_{n}y_{n}(z)\left[l_{2}K_{0}(\mu_{n}R)+\Lambda^{2}\mu_{n}K_{1}(\mu_{n}R)\right]=l_{2}.\label{eq:42}
\end{equation}
Multiplying this equation with $y_{n}(z)$ and integrating over $z$,
we find 
\begin{equation}
C_{n}=\frac{l_{2}p_{n}}{q_{n}}\frac{1-\cos(q_{n}t)}{l_{2}K_{0}(\mu_{n}R)+\Lambda^{2}\mu_{n}K_{1}(\mu_{n}R)}.\label{eq:43}
\end{equation}

Thus, the total flux $f(r,z)$ of excitons to the dislocation from
a point source at ($r,z$) is given by Eq.~(\ref{eq:40}), where
the eigenvalues $q_{n}$ are solutions of Eq.~(\ref{eq:37}), the
functions $y_{n}(z)$ are given by Eq.~(\ref{eq:36}) and the coefficients
$C_{n}$ by Eq.~(\ref{eq:43}).

\subsubsection{Flux to the planar boundary}

The flux to the planar surface $g(r,z)$ due to a point source at
$(r,z)$ is given, by virtue of the reciprocity theorem, by a solution
of the equation 
\begin{equation}
\Delta_{r}g(r,z)+\frac{\partial^{2}g(r,z)}{\partial z^{2}}-\Lambda^{-2}g(r,z)=0\label{eq:44}
\end{equation}
with the boundary conditions 
\begin{gather}
\left.\left(-\Lambda^{2}\frac{\partial g}{\partial z}+l_{1}g\right)\right|_{z=0}=l_{1},\,\,\,\left.\left(-\Lambda^{2}\frac{\partial g}{\partial r}+l_{2}g\right)\right|_{r=R}=0,\nonumber \\
g(r,t)=0.\label{eq:45}
\end{gather}

Similarly to Sec.~\ref{subsec:AbsorbPlanar}, we seek first for a
function $v(z)$ solving Eq.~(\ref{eq:44}) with the boundary conditions
(\ref{eq:45}) at $z=0$ and $z=t$, i.\,e., 
\begin{equation}
\frac{d^{2}v}{dz^{2}}-\Lambda^{-2}g=0,\,\,\,\left.\left(-\Lambda^{2}\frac{dv}{dz}+l_{1}v\right)\right|_{z=0}=l_{1},\,\,\, v(t)=0.\label{eq:46}
\end{equation}
The solution is 
\begin{equation}
v(z)=\frac{l_{1}\sinh[(t-z)/\Lambda]}{\Lambda\cosh(t/\Lambda)+l_{1}\sinh(t/\Lambda)}.\label{eq:47}
\end{equation}

Now we introduce a new function $u(r,z)$ with 
\begin{equation}
g(r,z)=v(z)-u(r,z).\label{eq:47a}
\end{equation}
The boundary value problem for the function $u(r,z)$ reads 
\begin{equation}
\Delta_{r}u(r,z)+\frac{\partial^{2}u(r,z)}{\partial z^{2}}-\Lambda^{-2}u(r,z)=0,\label{eq:48}
\end{equation}
\begin{gather}
\left.\left(-\Lambda^{2}\frac{\partial u}{\partial z}+l_{1}u\right)\right|_{z=0}=0,\nonumber \\
\left.\left(-\Lambda^{2}\frac{\partial u}{\partial r}+l_{2}u\right)\right|_{r=R}=l_{2}v(z),\,\,\, u(r,t)=0.\label{eq:49}
\end{gather}

We seek a solution in the same form as in Eq.~(\ref{eq:40}) above,
\begin{equation}
u(r,z)=\sum_{n=1}^{\infty}C_{n}y_{n}(z)K_{0}(\mu_{n}r),\label{eq:49a}
\end{equation}
and arrive at the equations for the coefficients $C_{n}$: 
\begin{equation}
\sum_{n=1}^{\infty}C_{n}y_{n}(z)\left[l_{2}K_{0}(\mu_{n}R)+\Lambda^{2}\mu_{n}K_{1}(\mu_{n}R)\right]=l_{2}v(z),\label{eq:50}
\end{equation}
that replace Eq.~(\ref{eq:42}) in the previous section. Multiplying
this equation by $y_{n}(z)$ and taking into account that the functions
$y_{n}(z)$ are orthogonal and normalized according to Eq.~(\ref{eq:38}),
we find the coefficients 
\begin{eqnarray}
C_{n} & = & \frac{l_{1}l_{2}\Lambda p_{n}}{\Lambda\cosh(t/\Lambda)+l_{1}\sinh(t/\Lambda)}\label{eq:51}\\
 & \times & \frac{\cosh(t/\Lambda)\sin(q_{n}t)-q_{n}\Lambda\sinh(t/\Lambda)\cos(q_{n}t)}{\left(1+q_{n}^{2}\Lambda^{2}\right)\left[l_{2}K_{0}(\mu_{n}R)+\Lambda^{2}\mu_{n}K_{1}(\mu_{n}R)\right]}.\nonumber 
\end{eqnarray}
~

Thus, the total flux $g(r,z)$ to the planar surface from a point
source at ($r,z$) is given by Eq.~(\ref{eq:47a}), where $v(z)$
is defined by Eq.~(\ref{eq:47}) and $u(r,z)$ by Eq.~(\ref{eq:49a}),
with the coefficients $C_{n}$ given by Eq.~(\ref{eq:51}). The function
$g(r,z)$ directly gives the EBIC signal, while the CL intensity is
equal to $1-\left[f(r,z)+g(r,z)\right]$, where the function $f(r,z)$
is given by Eq.~(\ref{eq:40}).

\subsubsection{Finite recombination velocity at the back surface}

A finite exciton recombination velocity $S_{3}$ at the back surface
$z=t$ gives rise to a surface recombination length $l_{3}=S_{3}\tau$,
analogously to the recombination length $l_{1}$ at the planar surface
$z=0$ and the recombination length $l_{2}$ at the dislocation $r=R$.
Then, the boundary condition (\ref{eq:33})$_{3}$ for the diffusion
equation (\ref{eq:32}) is replaced with 
\begin{equation}
\left.\left(\Lambda^{2}\frac{\partial f}{\partial z}+l_{3}f\right)\right|_{z=t}=0,\label{eq:52}
\end{equation}
and the boundary condition (\ref{eq:35})$_{2}$ for the eigenfunctions
$y_{n}(z)$ becomes 
\begin{equation}
\Lambda^{2}\left.\frac{dy_{n}}{dz}\right|_{z=t}+l_{3}y_{n}(t)=0.\label{eq:53}
\end{equation}

The solutions of Eq.~(\ref{eq:34}) with boundary conditions (\ref{eq:35})$_{1}$
and (\ref{eq:53}) are 
\begin{equation}
y_{n}(z)=p_{n}\left[\sin(q_{n}z)+\left(\Lambda^{2}q_{n}/l_{1}\right)\cos(q_{n}z)\right],\label{eq:54}
\end{equation}
where the eigenvalues $q_{n}$ are the solutions of the equation 
\begin{equation}
\tan(q_{n}t)=\frac{(l_{1}+l_{3})q_{n}}{\Lambda^{2}q_{n}^{2}-l_{1}l_{3}/\Lambda^{2}}\label{eq:55}
\end{equation}
and the coefficients $p_{n}$ providing the normalization condition
(\ref{eq:38}) are given by 
\begin{equation}
p_{n}^{2}=\left\{ \frac{l_{1}^{2}+q_{n}^{2}\Lambda^{4}}{2l_{1}^{2}}\left[t+\frac{\Lambda^{2}(l_{1}+l_{3})(l_{1}l_{3}+q_{n}^{2}\Lambda^{4})}{(l_{1}^{2}+q_{n}^{2}\Lambda^{4})(l_{3}^{2}+q_{n}^{2}\Lambda^{4})}\right]\right\} ^{-1}.\label{eq:56}
\end{equation}

The functions $y_{n}(z)$ defined by Eq.~(\ref{eq:54}) can be used
in Eqs.~(\ref{eq:40}) and (\ref{eq:49a}) to obtain the fluxes to
the dislocation and to the upper planar boundary, respectively.

\section{Examples}

\subsection{CL profile in the absence of surface recombination}

Let us analyze first the most simple case for which the recombination
of excitons at the planar surface is absent. The CL intensity is equal
to $I_{\mathrm{CL}}(r)=1-f(r)$, where $f(r)$ is given by Eq.~(\ref{eq:14}).
Since the radius of the dislocation $R$ is always small compared
to the diffusion length $\Lambda$, we can use the expansion of the
Bessel function for small arguments, $K_{1}(x)\approx1/x$, and represent
the CL intensity as 
\begin{equation}
I_{\mathrm{CL}}(r)=1-\frac{K_{0}(r/\Lambda)}{K_{0}(R/\Lambda)+\Lambda^{2}/Rl_{2}}.\label{eq:57}
\end{equation}

If the dislocation is a perfect sink for excitons, $l_{2}\rightarrow\infty$,
and Eq.~(\ref{eq:57}) reduces to 
\begin{equation}
I_{\mathrm{CL}}(r)=1-\frac{K_{0}(r/\Lambda)}{K_{0}(R/\Lambda)}.\label{eq:58}
\end{equation}
This expression was derived first by Brantley \emph{et al}. \cite{brantley75}
for spatially uniform generation of carriers and local detection,
whereas we consider local (point source) generation and integral detection.
At the first glance, this coincidence is surprising. However, it follows
directly from the reciprocity theorem proven in Sec.~\ref{sec:Reciprocity}:
the problems are actually exactly reciprocal, so their solutions coincide.
In the absence of surface recombination, the diffusion problems in
half-space and in infinite space also coincide, since the trajectory
of the diffusional motion of an exciton reflected from the surface
of the half-space is a mirror image of the continuation of that random
trajectory in the other half-space. Hence, the solution of the point-generation
problem in the half-space in the absence of surface recombination
and with the dislocation as a perfect sink is given by Eq.~(\ref{eq:58}).
The generalization to the case of finite recombination velocities
at the surface and at the dislocation is derived in Sec.~\ref{sec:Calculation}.

Suzuki and Matsumoto \cite{suzuki75} as well as Lax \cite{lax78}
noted that Eq.~(\ref{eq:58}) may, \emph{for large arguments}, be
approximated by replacing the Bessel function by its asymptotic 
\begin{equation}
I_{\mathrm{CL}}(r)=1-\exp(-r/\Lambda).\label{eq:59}
\end{equation}
This approximation was used in essentially all studies aiming to derive
the exciton diffusion length in GaN from CL experiments \cite{sugahara98,cherns01,pauc06,pauc10,nakaij05,ino08}.
While it may be justified to use Eq.~(\ref{eq:58}) in the case when
surface recombination can be neglected, the replacement of the Bessel
function by Eq.~(\ref{eq:59}) is a valid approximation only at $r\gg\Lambda$,
where the CL contrast vanishes. In fact, a reliable determination
of the diffusion length requires the accurate description of the profile
at $r\lesssim\Lambda$, which is not possible with Eq.~(\ref{eq:59}).

\begin{figure}
\includegraphics[width=0.9\columnwidth]{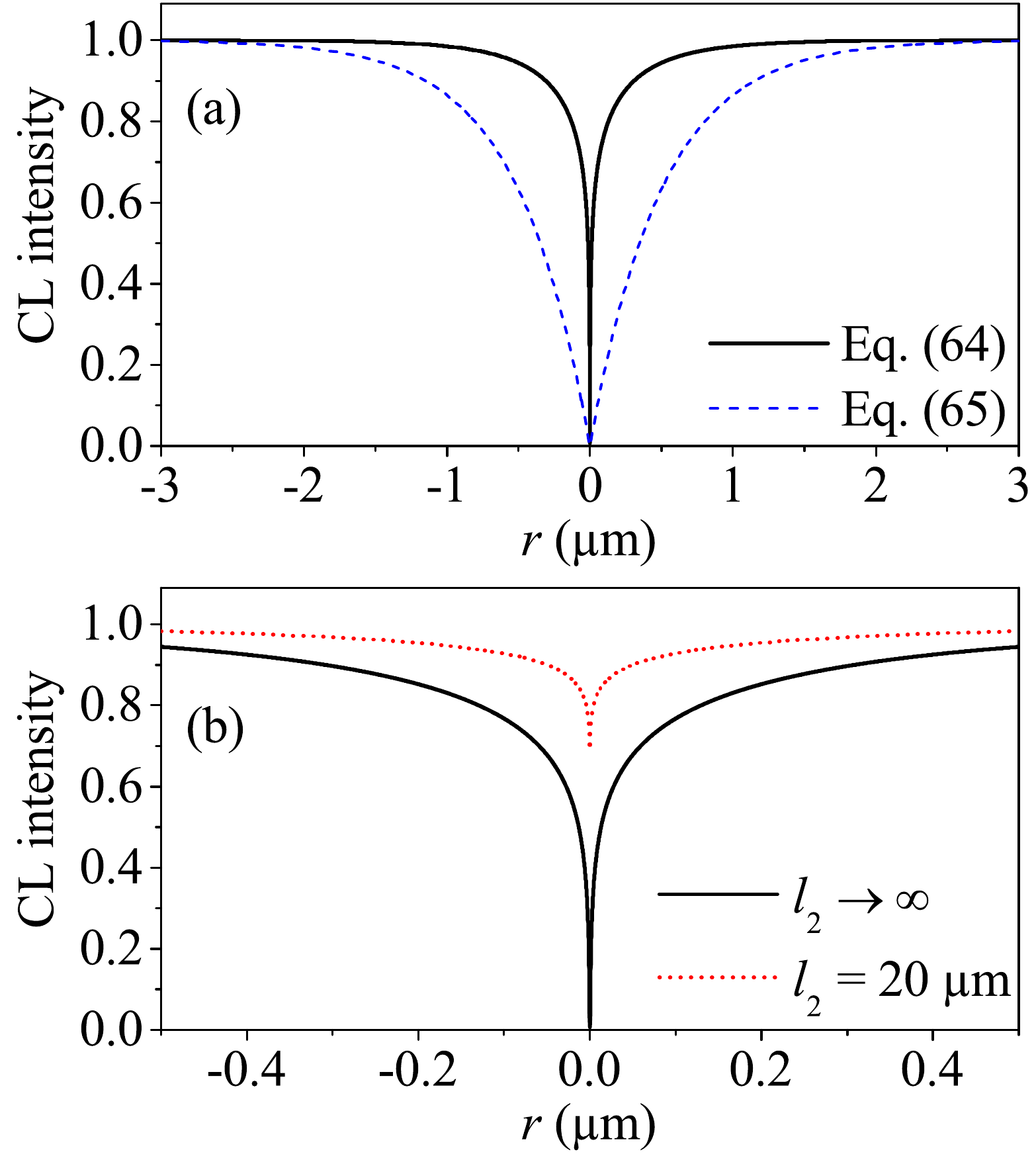}

\protect\protect\caption{Calculated CL contrast of a dislocation in the absence of surface
recombination at the planar surface ($l_{1}=0$). An exciton diffusion
length of $\Lambda=0.5$~\textmu m and a dislocation radius of $R=1$~nm
are assumed. (a) Infinite recombination velocity at the dislocation
line $(l_{2}\rightarrow\infty)$: Comparison of the exact solution
Eq.~(\ref{eq:58}) (solid line) and its exponential asymptotic approximation
Eq.~(\ref{eq:59}) (dashed line). (b) A comparison of infinite {[}$l_{2}\rightarrow\infty$,
Eq.~(\ref{eq:58}), thick line{]} and finite {[}$l_{2}=$20~\textmu m,
Eq.~(\ref{eq:57}), dotted line{]} recombination velocities at the
dislocation.}

\label{fig:ReflCond} 
\end{figure}

Figure \ref{fig:ReflCond}(a) compares the dislocation contrast given
by Eq.~(\ref{eq:58}) with its common replacement \cite{rosner97,sugahara98,cherns01,pauc06,pauc10,nakaij05,ino08}
by an exponential function {[}Eq.~(\ref{eq:59}){]} for the case
when the dislocation acts as a perfect sink for excitons. The exciton
diffusion length is assumed to be $\Lambda=$0.5~\textmu m. The dislocation
radius for the exciton diffusion problem cannot be determined from
experiments, and we assume it to be of atomic scale, $R=$1~nm. Since
the ratio $R/\Lambda$ is certainly small and the Bessel function
$K_{0}(R/\Lambda)$ depends on its argument only logarithmically for
small arguments, the choice of $R$ does not play a role when the
dislocation acts as a perfect sink for the excitons ($l_{2}\rightarrow\infty$).
It is evident from Fig.~\ref{fig:ReflCond}(a) that the replacement
of the solution (\ref{eq:58}) (thick line) with the exponential function
(\ref{eq:59}) (dashed line), as it is done in Refs.~\cite{rosner97,sugahara98,cherns01,pauc06,pauc10,nakaij05,ino08},
is an exceedingly poor approximation which leads to a large error
in the determination of the exciton diffusion length from the CL contrast
of dislocations.

Figure \ref{fig:ReflCond}(b) shows the effect of a finite recombination
velocity at the dislocation (finite $l_{2}$). The CL intensity depends
on the product $l_{2}R$, as it follows from Eq.~(\ref{eq:57}).
Since both parameters characterizing a dislocation, $R$ and $l_{2}$,
are difficult to determine separately from an experiment, we keep
a constant value of $R=$1~nm in the calculations below and vary
$l_{2}$ to match the contrast. The decrease of this product results
in a decrease of the contrast, as it is shown by the dotted line in
Fig.~\ref{fig:ReflCond}(b).

\subsection{The effect of surface recombination}

\begin{figure}
\includegraphics[width=0.9\columnwidth]{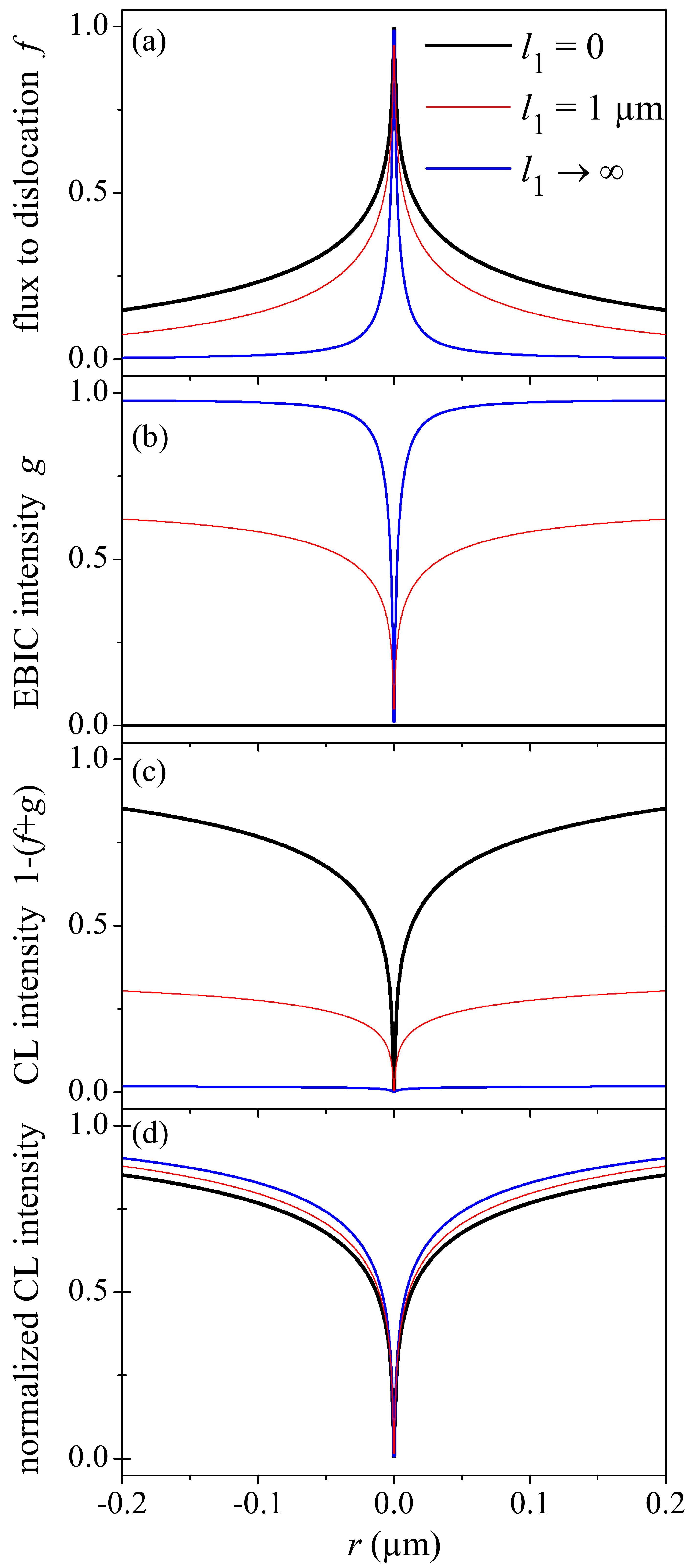}

\protect\protect\caption{(a) Total flux of excitons to the dislocation $f$, (b) total flux
to the planar surface $g$ (the EBIC signal), (c) fraction of excitons
experiencing radiative recombination in the bulk $1-(f+g)$ (the CL
intensity), and (d) CL intensity scaled by the intensity far from
the dislocation. A point source of excitons is located at a distance
$r$ from the dislocation and at a depth $z=10$~nm from the surface.
A diffusion length of $\Lambda=0.5$~\textmu m is assumed. The dislocation
is considered to be a perfect sink for excitons ($l_{2}\rightarrow\infty$),
and the surface recombination velocity is zero ($l_{1}=0$, black
lines), finite ($l_{1}=1$~\textmu m, red lines), or infinite ($l_{1}\rightarrow\infty$,
blue lines).}

\label{fig:Absorb} 
\end{figure}

Figure \ref{fig:Absorb} presents the calculated dislocation contrast
for a point source of excitons located at a distance $r$ from the
dislocation line and at a depth $z=10$~nm from the surface. In these
calculations, we assume that the dislocation line is a perfect sink
for the excitons ($l_{2}\rightarrow\infty$). We use the formulas
derived above to calculate the total exciton flux to the dislocation
$f$ and their total flux to the planar surface $g$ (the EBIC signal
in an appropriately designed experiment). The remaining quantity $1-(f+g)$
accounts for the fraction of excitons experiencing recombination in
the bulk. As we have already discussed in Sec.~\ref{sec:equations},
only a fraction $\eta=\tau/\tau_{r}$ of excitons recombines radiatively
and thus contributes to the experimental observable, i.\,e., the
CL intensity. However, since the CL intensity cannot be measured on
an absolute scale, this factor can be omitted in the analysis.

One can see from Fig.~\ref{fig:Absorb}(a) that the exciton flux
to the dislocation $f$ is maximum when recombination at the planar
surface is absent and minimum when the planar surface acts as a perfect
sink for excitons reaching it. In turn, the flux to the planar surface
in Fig.~\ref{fig:Absorb}(b) is absent when the surface recombination
is absent and maximum when the surface acts as a perfect absorber.
In this latter case, as it is seen by the blue line in Fig.~\ref{fig:Absorb}(c),
the remaining fraction of excitons that are radiatively recombining
in the bulk and actually contribute to the CL signal is small. Far
from the dislocation line, only 2\% of the excitons are recombining
radiatively, since the source is close to the surface.

Since the CL intensities cannot be compared on the absolute scale,
they need to be normalized by calculating the contrast $(I-I_{\infty})/I_{\infty}$,
where $I_{\infty}$ is the intensity far from the dislocation. Then
all CL curves tend to 1 in the limit $r\rightarrow\infty$. The result
of this normalization is presented in Fig.~\ref{fig:Absorb}(d),
and it is seen that the normalized intensity only weakly depends on
the surface recombination velocity. Note that the calculations in
Fig.~\ref{fig:Absorb} are performed assuming that the dislocation
line is a perfect sink for the excitons. If the dislocation possesses
a finite recombination velocity (finite $l_{2}$), the contrast decreases,
in the same way as it is shown by the dotted line in Fig.~\ref{fig:ReflCond}(b).

\subsection{Finite source size\label{subsec:FiniteSource}}

\begin{figure}[b]
\includegraphics[width=0.9\columnwidth]{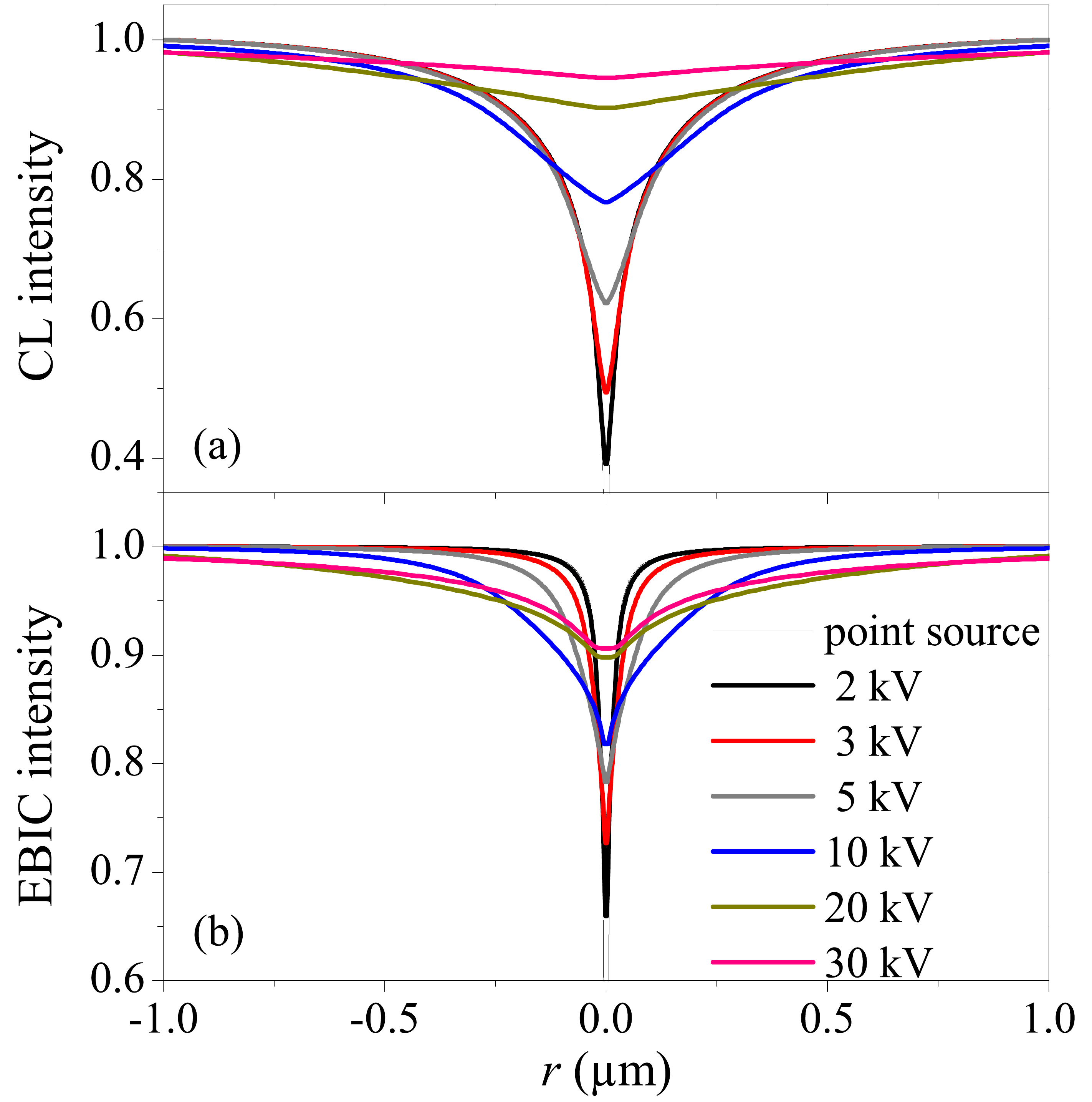}

\protect\protect\caption{Calculated (a) CL and (b) EBIC contrast of a dislocation for different
acceleration voltages of the electron beam. For the calucaltions,
we assumed total annihilation of excitons at both the planar surface
and at the dislocation (i.e., $l_{1}\rightarrow\infty,\,\, l_{2}\rightarrow\infty$)
and an exciton diffusion length of $\Lambda=0.5$~\textmu m.}

\label{fig:source} 
\end{figure}

The solutions obtained in Sec.~\ref{sec:Calculation} are the Green
functions that describe the CL and EBIC signals due to a point source
of excitons located at a distance $r$ from the dislocation line and
at a depth $z$. Since the diffusion problem is linear, the solution
for an arbitrary finite source of excitons can be obtained as the
convolution integral (\ref{eq:3a}) of the Green function with the
initial exciton distribution produced by the electron beam.

We employ the common assumption \cite{toth98} that the number of
thermalized electron-hole pairs (and hence excitons) produced in a
volume element by the electron beam is proportional to the energy
loss of the electrons in that volume, and use the Monte Carlo software
$\mathtt{CASINO}$ \cite{demers11,demers12} to simulate the distributions
of the energy losses at different acceleration voltages of the electron
beam. We do not attempt to describe the initial exciton distribution
by any analytical function, as it is done in other works \cite{donolato81,donolato82,werner88,cherns01,parish06,yakimov07}.
Rather, we simulate the initial exciton distribution by $\mathtt{CASINO}$
on a $100\times100\times100$ grid and perform a direct summation
of intensities considering each point of the grid as a point source,
with the number of excitons in it proportional to the local energy
loss of the electron beam.

Figures \ref{fig:source}(a) and \ref{fig:source}(b) compare, respectively,
the CL and EBIC profiles at different acceleration voltages, assuming
that both the planar surface and the dislocation are perfect sinks
for the excitons ($l_{1}\rightarrow\infty,\,\, l_{2}\rightarrow\infty$).
One can see that, for low acceleration voltages up to 3~kV, the profiles
are affected only in a very narrow range near the dislocation line.
As the acceleration voltage is beyond 3~kV, however, the profiles
broaden significantly and the contrast decreases. The CL profiles
are noticeably broader than the EBIC profiles. The widths of the CL
profiles monotonously increase with the acceleration voltage, while
the EBIC profiles saturate at high acceleration voltages, since the
excitons that are generated at depths exceeding the diffusion length
do not reach the surface \cite{shmidt02}.

\subsection{Experimental profiles}

\begin{figure}
\includegraphics[width=0.8\columnwidth]{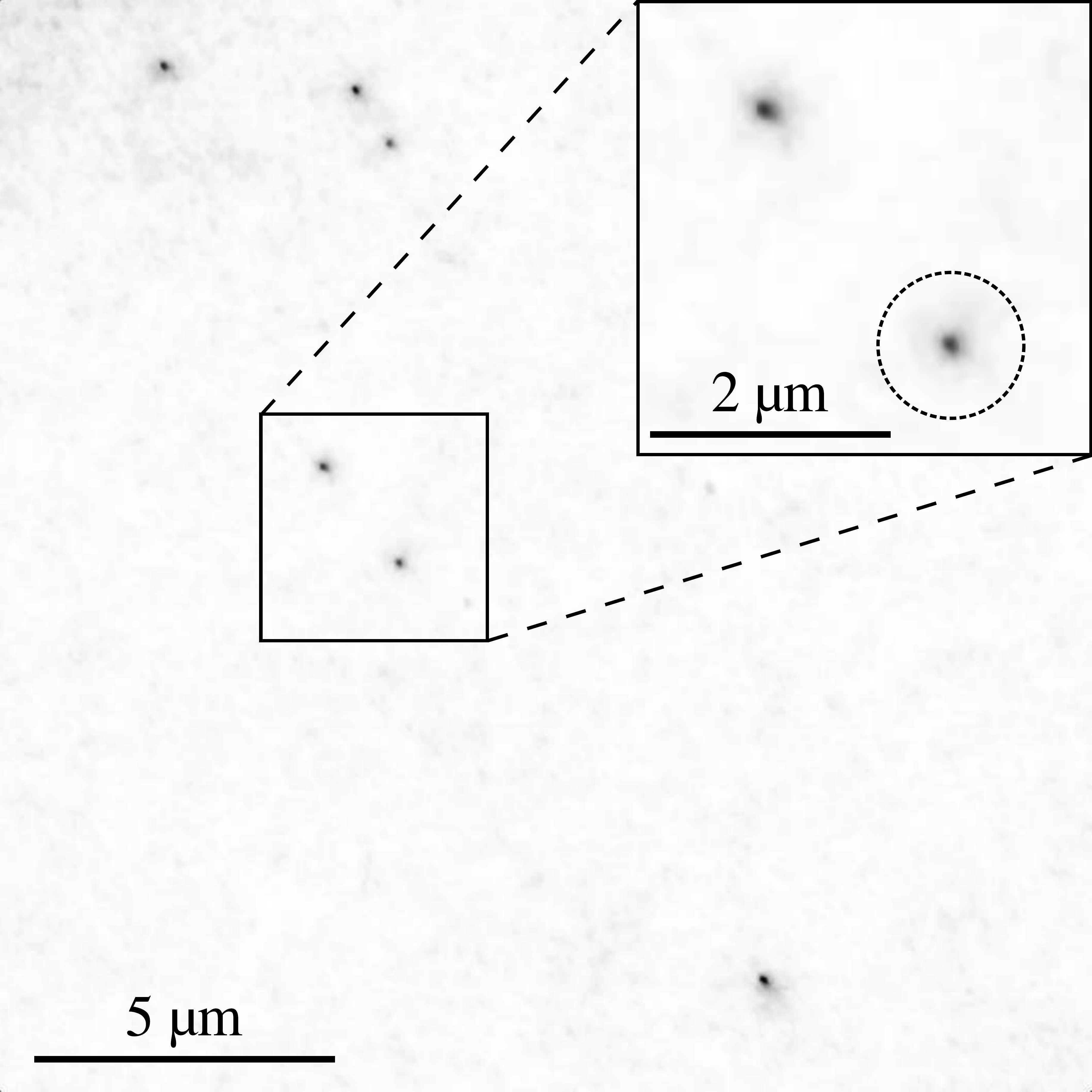}

\protect\caption{Panchromatic CL map of the free-standing GaN(0001) layer under investigation
over an area of 330~\textmu m$^{2}$. The CL intensity is essentially
uniform except at the outcrops of threading dislocations, where clearly
defined dark spots are observed. The inset shows a magnified view
of two threading dislocations. For further analysis, a linescan was
taken across the dislocation marked with a dashed circle (see Fig.\ref{fig:CLprofile}).}

\label{fig:CLimage} 
\end{figure}

CL spectroscopy was carried out in a Zeiss Ultra55 field-emission
scanning electron microscope equipped with a Gatan MonoCL4 system
and a He-cooling stage. A photomultiplier tube was used for the acquisition
of panchromatic CL images. The acceleration voltage was set to 3~kV,
and the probe current of the electron beam to 0.75~nA, with the aim
to minimize the generation volume of excitons as much as possible
{[}cf.\,Fig.\,\ref{fig:source}(a){]}.

Figure \ref{fig:CLimage} presents a panchromatic CL map of a $\approx350$~\textmu m
thick free-standing GaN(0001) layer grown by hydride vapor phase epitaxy.
The average threading dislocation density present at the surface of
this layer amounts to about $6\times10^{5}$~cm$^{-2}$, low enough
to allow us to perform intensity line scans of individual dislocations
not influenced by other ones. Furthermore, the exciton diffusion length
in this sample is certainly not limited by the average distance between
dislocations.

\begin{figure}[b]
\includegraphics[width=0.9\columnwidth]{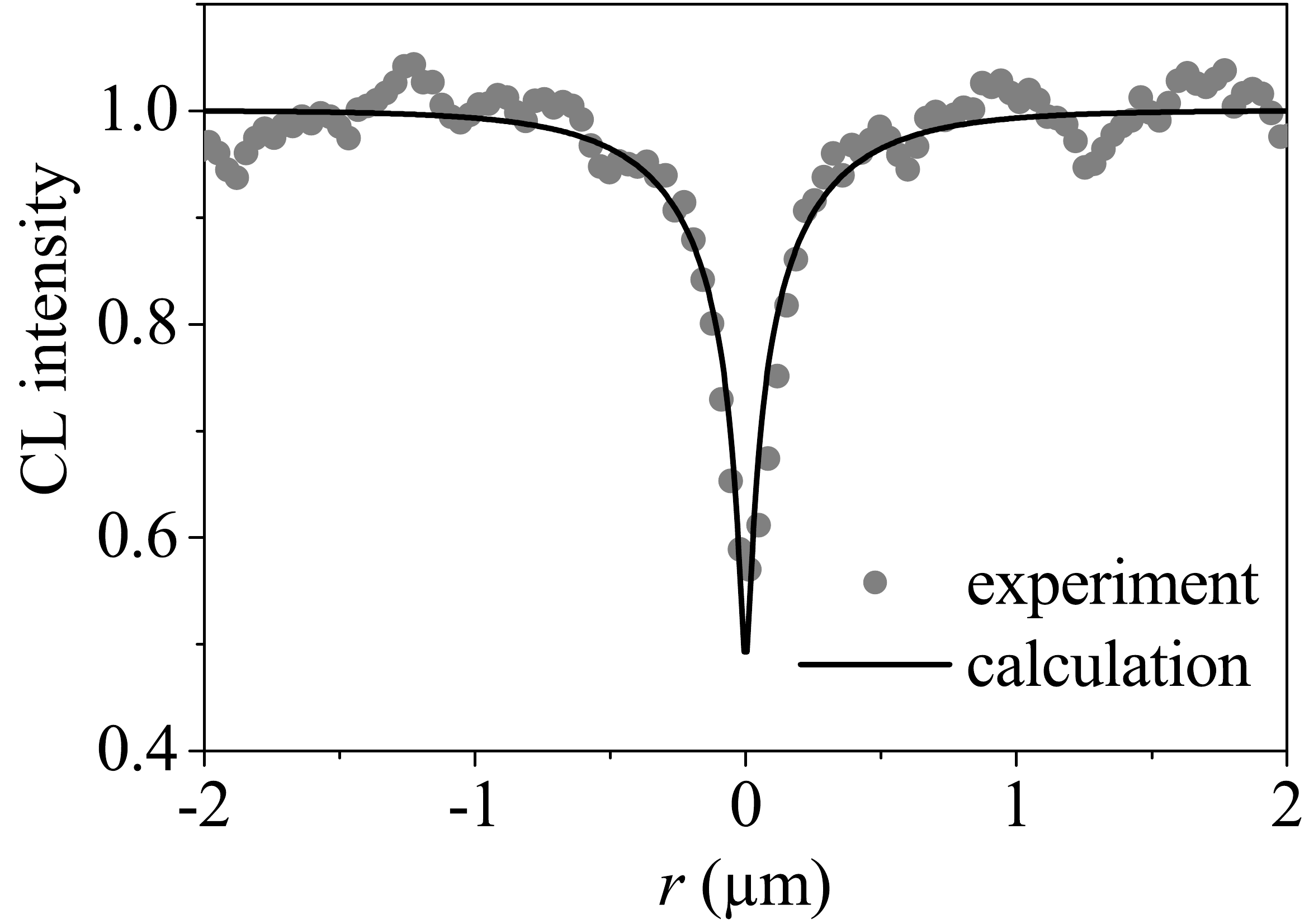}

\protect\protect\caption{Experimental CL intensity profile across a threading dislocation in
our free-standing GaN layer (symbols) together with a calculated profile
(line) obtained by integrating over the simulated exciton distribution
generated by the electron beam with an acceleration voltage of 3\,keV,
with an exciton diffusion length of $\Lambda=0.4$~\textmu m, a surface
recombination length of $l_{1}=0.5$~\textmu m, a dislocation radius
$R=1$~nm, and a dislocation recombination length of $l_{2}=275$~\textmu m.}

\label{fig:CLprofile} 
\end{figure}

Figure \ref{fig:CLprofile} compares an experimental intensity profile
with the calculated one based on the analytical expressions derived
in Sec.~\ref{subsec:Finite} and a summation over the initial spatial
distribution of the excitons, simulated by the calculated energy loss
distribution of the incident electron beam. As we have already seen
in Fig.~\ref{fig:Absorb}(d), the CL intensity profile, after it
is scaled such that the intensity far from the dislocation is unity,
is fairly insensitive to the recombination velocity at the planar
boundary. Hence, we estimate the surface recombination length using
the results of an independent experimental determination of the surface
recombination velocity and the exciton lifetime in GaN epilayers grown
by metalorganic chemical-vapor deposition \cite{aleksieunas03}. This
latter study provides values of $D=1.7$~cm$^{2}$/s, $S=5\times10^{4}$~cm/s,
and $\tau\approx1$~ns, which gives $\Lambda=\sqrt{D\tau}=0.4$~\textmu m
and $l_{1}=S\tau=0.5$~\textmu m.

We fit the experimental curve in Fig.~\ref{fig:CLprofile} taking
the surface recombination length fixed at $l_{1}=0.5$~\textmu m
and obtain a diffusion length $\Lambda=0.4$~\textmu m, in an excellent
agreement with the results of Ref.~\cite{aleksieunas03}, and the
recombination length at the dislocation $l_{2}=275$~\textmu m. When
we fix the surface recombination length $l_{1}$ at values either
larger or smaller by an order of magnitude, we obtain fits practically
coinciding with the one shown in Fig.~\ref{fig:CLprofile}. The values
of the exciton diffusion length and the dislocation recombination
length are found to be $\Lambda=0.52$~\textmu m and $l_{2}=356$~\textmu m
in the former and $\Lambda=0.35$~\textmu m and $l_{2}=148$~\textmu m
in the latter case, respectively. These values provide an estimate
of the accuracy in determination of the parameters in the case when
the surface recombination length is not known.

\section{Summary and Conclusions}

We have presented a rigorous, complete, and general solution for the
problem of excitons with an arbitrary initial distribution recombining
in the bulk as well as diffusing toward a planar (the surface) and
a linear (the dislocation) defect both being characterized by arbitrary,
finite recombination velocities. This solution gives access, in principle,
to three technologically important quantities: the surface recombination
velocity, the recombination strength at the dislocation, and the exciton
diffusion length. Our calculations have shown, however, that the former
of these quantities has only a minor impact as long as the CL intensity
cannot be measured on an absolute scale. As a result, normalized intensity
profiles allow one to uniquely extract both the recombination strength
at the dislocation and the exciton diffusion length.

Regarding the former, our analysis shows that the threading dislocations
in GaN(0001) layers are extremely efficient nonradiative centers,
similar to those in other III-V compounds \cite{hildebrandt98}. Concerning
the latter, let us first stress that the diffusion length is not a
material constant, as it seems to be often perceived, but a figure
of merit for the purity and perfection of a given sample. More specifically,
the diffusion length depends on the exciton diffusivity, and thus,
via the Einstein relation, on its mobility, as well as on its lifetime.
At room temperature, the exciton mobility should be limited by polar
optical phonon scattering, and is thus expected to not vary significantly
among samples with moderate defect densities. However, the exciton
lifetime in GaN is still far from being determined by intrinsic radiative
recombination, but is controlled by nonradiative, Shockley-Read-Hall
type recombination at defect states \cite{scajev12}. The room temperature
diffusion length in samples with very low dislocation density, as
the one used in the present work, thus provides insight into the impact
of point defects on the internal quantum efficiency of GaN.

\section{Acknowledgments}

We are indebted to Uwe Jahn for a critical reading of the manuscript.
K. K. S. kindly acknowledges the support of Russian Science Foundation
under grant N 14-11-00083.

%\bibliographystyle{aipnum4-1}\bibliographystyle{plain}\bibliographystyle{plain}
% \bibliography{CL}

%merlin.mbs apsrev4-1.bst 2010-07-25 4.21a (PWD, AO, DPC) hacked
%Control: key (0)
%Control: author (0) dotless jnrlst
%Control: editor formatted (1) identically to author
%Control: production of article title (0) allowed
%Control: page (1) range
%Control: year (0) verbatim
%Control: production of eprint (0) enabled
%

\end{document}